\documentclass[twocolumn,showpacs,prb,preprintnumbers,amsmath,amssymb]{revtex4}
\usepackage[dvips]{graphicx}
\usepackage{dcolumn}
\usepackage{bm}

\begin{document}

\title{Study of superconducting properties of MgB$_{2}$}

\author{Y. Machida}
 \altaffiliation{Corresponding author: Tel.+81-45-924-5383; Fax.+81-45-924-5339\\
E-mail address: machida@lipro.rlem.titech.ac.jp}
\author{S. Sasaki}
\affiliation{Materials and Structures Laboratory, Tokyo Institute of Technology
\\4259 Nagatsuta, Midori-ku, Yokohama 226-8503, Japan}

\author{H. Fujii}
\affiliation{National Institute for Materials Science, 1-2-1, Sengen,Tsukuba 305-0047, Japan}

\author{M. Furuyama, I. Kakeya and K. Kadowaki}
\affiliation{Institute of Materials Science, University of Tsukuba,1-1-1, Tennoudai,
 Tsukuba 305-8573, Japan}


\date{\today}

\begin{abstract}
We synthesized single crystalline and policrystalline MgB$_{2}$ under ambient pressures. 
The single crystals of MgB$_{2}$ were of good quality, where the crystal structure refinements 
were successfully converged with \textit{R}\,=\,0.020.
The specific heat of policrystalline MgB$_{2}$ samples has been measured in a temperature range between 2 and 60\,K 
in magnetic field up to 6\,T. 
The measurement gave the coefficient of the linear term in the electronic specific heat,  
$\gamma$\,=\,3.51\,mJ/K$^{2}$\,mol, and the jump of the specific heat, 2.8\,mJ/K$^{2}$\,mol at 38.5\,K. 
It is shown from the analysis of the specific heat that the electronic specific heat 
in the superconducting state differs largely from the conventional BCS weak coupling theory.
From the results of measurements of the magnetic properties on single crystal samples, 
we found a sharp superconducting
transition at 38\,K with transition width $\Delta$\textit{Tc}\,=0\,.8\,K and the superconducting
anisotropy ratio $\gamma$ increasing from about 1 near \textit{T}c to 4.0 at 25\,K.
\end{abstract}

\maketitle

The recent discovery of superconductivity at 39\,K in magnesium diboride MgB$_{2}$ \cite{nagamatsu} 
has attracted great scientific interest. Several experiments indicated a phonon-mediated \textit{s}-wave BCS 
superconductivity\cite{budko,quilty} and the appearance of a double energy gap was predicted \cite{kotus,liu}. 
Specific heat \cite{bouquet}
and spectroscopic \cite{szabo}
 measurements, scanning tunneling spectroscopy \cite{giubileo,iavarone} gave evidence for this prediction. 
However, several key parameters such as the upper critical fields \textit{H}$_{c2}$ and their 
anisotropy ratio $\gamma$, the magnetic penetration $\lambda$, 
the coherence lengths $\xi$ and Ginzburg-Landau parameters $\kappa$ are not well 
established because of the difficulty of growing high quality MgB$_{2}$ single crystals. Especially the anisotropy ratio 
$\gamma$=\textit{H}$_{c2}^{ab}$/\textit{H}$_{c2}^{c}$ is important to clarify the superconducting mechanism and applications of 
MgB$_{2}$. Here \textit{H}$_{c2}^{ab}$ and \textit{H}$_{c2}^{c}$
 are the in-plane and the out-of-plane upper critical fields respectively. Reported $\gamma$-values vary widely depending on 
the measurement methods or on the sample types. The values determined from resistivity on pollycrystallie \cite{simon}, 
aligned crystallites \cite{lima}, c-axis oriented films \cite{patnaik,jung2,olsson} and 
single crystals \cite{lee,pradhan,kim,xu,angst,welp,zehetmayer,sologubenko2} have been reported to be 
6-9, 1.7, 1.3-2 and 2.6-3, respectively.

In this paper, we present a study of the specific and the magnetization measurements on single crystalline 
and policrystalline MgB$_{2}$. 
The polycrystals for the specific measurement were synthesized as follows. 
The Mg (99.99\,$\%$, Furuya Metal Co.) ingot and 
the B (99.9\,$\%$, Furuuchi Chemical Co.) powder were pressed into a cylinder with a diameter of 13\,mm and a length of 
15\,mm. This was put in the BN crucible with a lid, and this crucible was encapsulated in a stainless (SUS304) 
tube in argon atmosphere.
 Then, this was reacted for three hours at 1100\,$^{\circ}$C in an electric furnace. The sample after reaction was a form
 of sintered porous lump. This was not so hard that the sample was cut from this lump in a shape of thin, square plate 
for the specific heat measurement. 

The single crystals were grown in the stainless (SUS304) tube. 
The SUS304 tube had an outside diameter of 32\,mm, 
a wall thickness of 1.5\,mm, and a length of 110\,mm. The inner surface of the tube was sealed by 
Mo sheet (99.95\,$\%$, Nilaco Co.) with the size of 0.05$\times$100$\times$200\,mm$^{3}$. One end 
of the SUS304 tube was pressed with a vise and sealed in 
an Ar gas atmosphere by arc welding. 
The starting materials of Mg chunk with the size of 3$\sim$5\,mm and B 
chunk which was cut out with the size 
of about 1\,cm$^{3}$ from B block was 
filled  inside of the tube. Then the other end of the tube was pressed with a vise and sealed in an Ar gas 
atmosphere by arc welding, as well. 
The samples were heated from room temperature to 1200\,$^{\circ}$C for 40 minutes and kept at 1200\,$^{\circ}$C 
for 12 hours, then slowly 
cooled to room temperature for 12 hours. The single crystals finally obtained were
about 100 $\sim$ 300\,$\mu$m, which had irregular shapes with shiny gold color when observed under a optical 
microscope.

\begin{figure}
\begin{center}
\includegraphics[width=12cm,height=12cm,keepaspectratio]{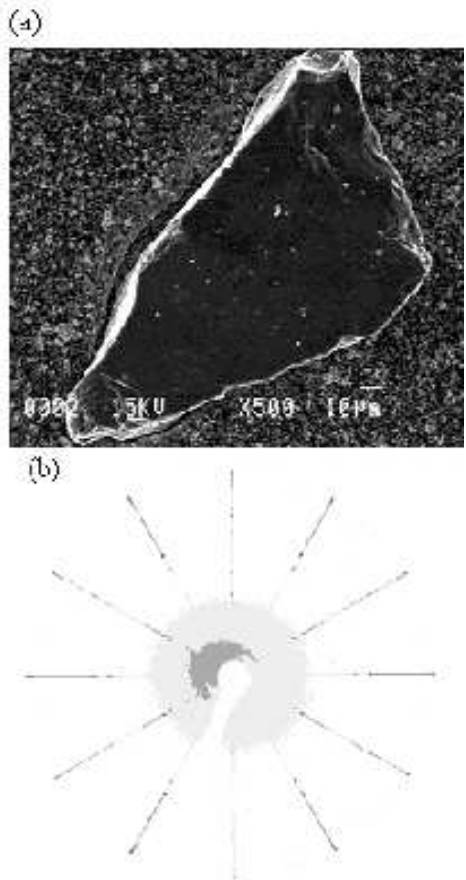}
\vspace*{-0.1cm}
\caption{(a)SEM image of a MgB$_{2}$ single crystal with a size of about 100\,$\mu$m. (b)X-ray precession photograph 
of the single crystal.}
	\label{fig:sem}
\end{center}
\end{figure}

\begin{figure}
\begin{center}
\includegraphics[width=8cm,height=8cm,keepaspectratio,clip]{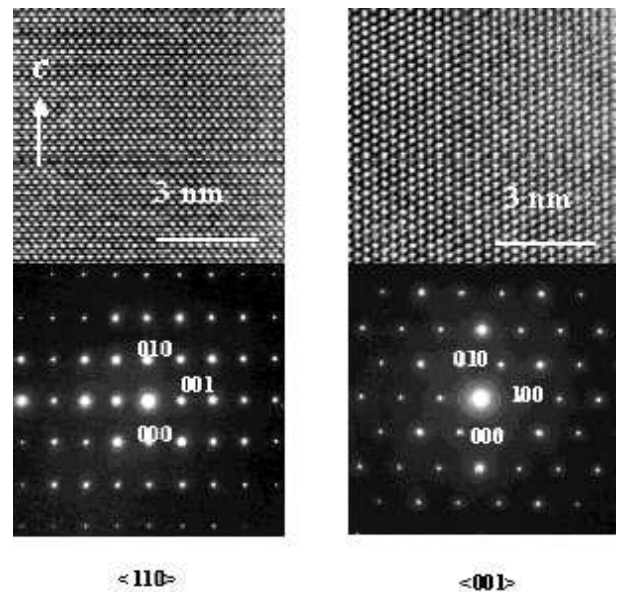}
\end{center}
\vspace*{-0.3cm}
	\caption{Electron diffraction patterns and HRTEM images of a MgB$_{2}$ single crystal 
for a beam direction of [001] and [110] in the hexagonal structure. }
	\label{fig:tem}
\end{figure}

The single crystal images observed by a scanning electron microscope (SEM) is shown in Fig \ref{fig:sem}(a). The crystals were found to have very flat surfaces. Structural analysis was carried out 
using a x-ray precession camera, a four-circle diffractometer and a transmission electron microscope (TEM). 
The x-ray precession photograph indicated that the crystal has the hexagonal structure, as shown in Fig \ref{fig:sem}(b).
The diffraction data were collected by using graphite monochromated Mo\textit{K}$_{\alpha}$ radiation at room 
temperature and refined by the least-square procedure using 86 reflections as the average of the measured 866 reflections.
 The obtained cell and structural parameters 
 were \textit{a}\,=\,\textit{b}\,=\,3.0863(4)\,\AA, \textit{c}\,=\,3.5178(4)\,\AA, 
(x,y,z)\,=\,(0,0,0)\,(Mg), (1/3,2/3,1/2)\,(B), 
\textit{U}$_{11}$(Mg)\,=\,0.0078(2)\,\AA$^{2}$, 
\textit{U}$_{33}$(Mg)\,=\,0.0054(2)\,\AA$^{2}$, \textit{B}$_{eq}$(Mg)\,=\,0.382(2)\,\AA$^{2}$,
 \textit{U}$_{11}$(B)\,=\,0.0068(2)\,\AA$^{2}$,  
\textit{U}$_{33}$(B)\,=\,0.0058(2)\,\AA$^{2}$ and \textit{B}$_{eq}$(B)\,=\,0.371(2)\,\AA$^{2}$ with small agreement factors 
\textit{R}\,=\,0.020, \textit{R}$_{w}$\,=\,0.027 (w\,=\,weight). 
To confirm the structure of the MgB$_{2}$ phase, we took plane-view HRTEM images and electron diffraction patterns
in selected areas for beam directions of [001] and [100] as shown in Fig \ref{fig:tem}, which indicated atomic arrangement 
with P6/mmm cell of MgB$_{2}$. Neither extra spots nor streaks were not found, indicating that the crystal is 
of high quality.

\begin{figure}
\begin{center}
\includegraphics[width=10cm,height=6.1cm,keepaspectratio]{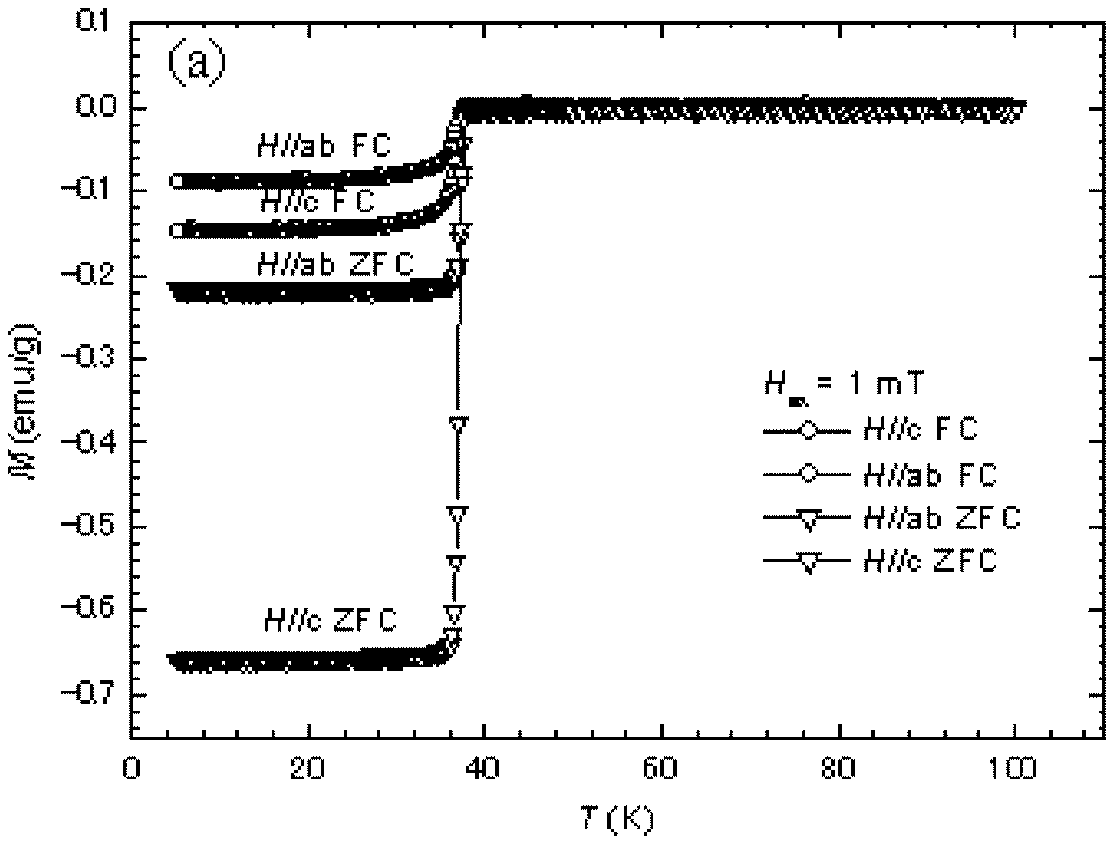}\\
\includegraphics[width=10cm,height=6cm,keepaspectratio]{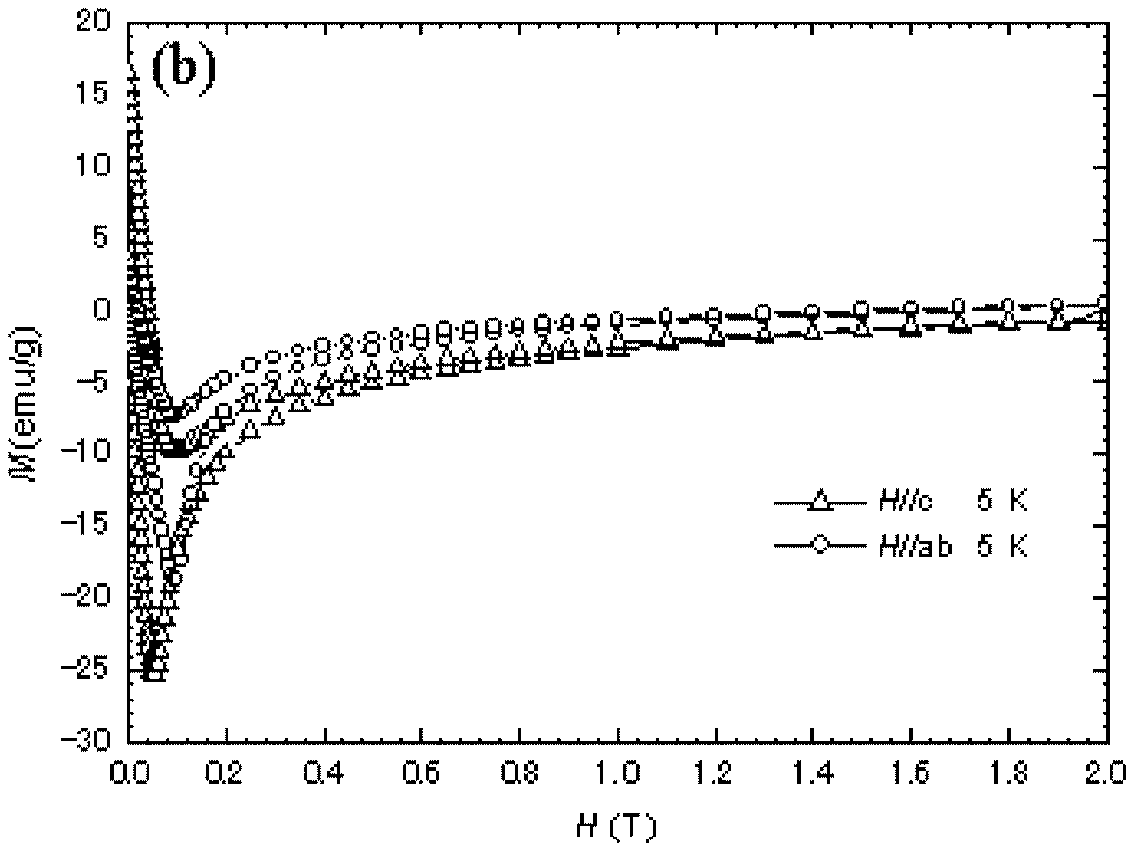}
\end{center}
\vspace*{-0.5cm}
	\caption{(a)Temperature dependent magnetization curves for the MgB$_{2}$ single crystal. 
FC and ZFC denote the field cooling 
and zero field cooling curves, respectively. (b)The magnetic hysteresis curves \textit{M}(\textit{H}) at 5\,K.}
	\label{fig:both}
  \label{fig:MH} 
\end{figure}

The temperature dependence of the magnetization 
curve was measured at 1\,mT along the \textit{c}-axis and the \textit{ab}-plane 
by a superconducting quantum interference device (SQUID) magnetometer. 
Figure \ref{fig:both}(a)  is the results 
for the MgB$_{2}$ single crystal. It shows the \textit{M}(\textit{T}) curves in the zero-field-cooling 
(ZFC) and the field-cooling (FC) modes. The onset of superconducting transition was observed at \textit{Tc}
\,=\,38\,K with transition width $\Delta$\textit{Tc}\,=\,0.8\,K both for \textit{H}//ab and for \textit{H}//c, indicating 
high quality 
of the samples, where //ab(//c) indicates the field \textit{H} perpendicular(parallel) 
to the \textit{c}-axis
 of the sample, respectively. Comparing to the transition temperatures ($\approx$39\,K) for polycrystal 
samples~\cite{jung,takano,jorgensen}, 
single crystal samples~\cite{lee,sologubenko,pradhan,kim,xu} formerly reported 
have a little bit lower transition temperature at around 38\,K. 
It is noted that our samples are not an exceptional case.  
This lower \textit{Tc} in single crystals was thought to be caused by contamination from container materials (BN, Mo, Nb).
From the quantitative analysis using an electron prove microanalyzer (EPMA), however we found that there was no contamination 
in the samples 
from the container, which was Mo and SUS304 in our case. According to this analysis,
 the composition in rather ideal without particular stoichiometry shift, therefore, 
the quality check of the single crystals is an urgent need in more detail.
 Figure \ref{fig:both}(b) shows the magnetic hysteresis curves \textit{M}(\textit{H}) 
at 5\,K for applied fields up to 
2\,T for \textit{H}//ab and for \textit{H}//c, indicating the characteristic curve of type-II superconductors. 
There is an asymmetry
between the ascending and the descending branches at \textit{H} $\leq$ 0.1\,T, which has been also observed in the 
magnetic hysteresis measurements on single crystals~\cite{zehetmayer}.

\begin{table*}
\caption{\label{tab:table}Comparison of physical parameters with single crystals prepared by different methods.}
\begin{ruledtabular}
\begin{tabular}{cccccc}
 Parameter&Our sample&J. Karpinski \textit{et al}~\cite{zehetmayer,angst,karpinski}&S. Lee \textit{et al}~\cite{lee,eltsev}
&M. Xu \textit{et al}~\cite{xu}&K. H. P. Kim \textit{et al}~\cite{kim,welp}\\ \hline
 \textit{a}(\AA)&3.0863(4)&3.085(1)&3.0851(5)&3.047(1)&3.09$\pm$0.06\\
 \textit{c}(\AA)&3.5178(4)&3.518(2)&3.5201(5)&3.404(1)&\\
 \textit{R}&0.020&0.015-0.020&0.018&&\\
 \textit{R}$_{w}$&0.027&0.015-0.020&0.025&&\\
 \textit{T}$_{c}$(K)&38&38-39&38.1-38.3&39&38\\\textit{H}$_{c2}^{ab}$(0)(T)&13.6&14.5\footnote{Values determined from the magnetic measurement},
23\footnote{Values determined from the magnetic torque measurement}
&21-22&19.8\\
 \textit{H}$_{c2}^{c}$(0)(T)&3.4&3.18\footnotemark[1],3.1\footnotemark[2]&7.0-7.5&7.7&3.5\\
 \textit{$\gamma$}(T)&1(\textit{T$_{c}$})-4.0(25\,K)&1(\textit{T$_{c}$})-4.2(22\,K)\footnotemark[1],
2.8(35\,K)-6(15\,K)\footnotemark[2]&
2.2(\textit{T$_{c}$})-3(30\,K)&2.6(0\,K)&2(\textit{T$_{c}$})-4.4(22\,K)
\end{tabular}
\end{ruledtabular}
\end{table*}

\begin{figure}
\begin{center}
\includegraphics[width=8cm,height=12cm,keepaspectratio]{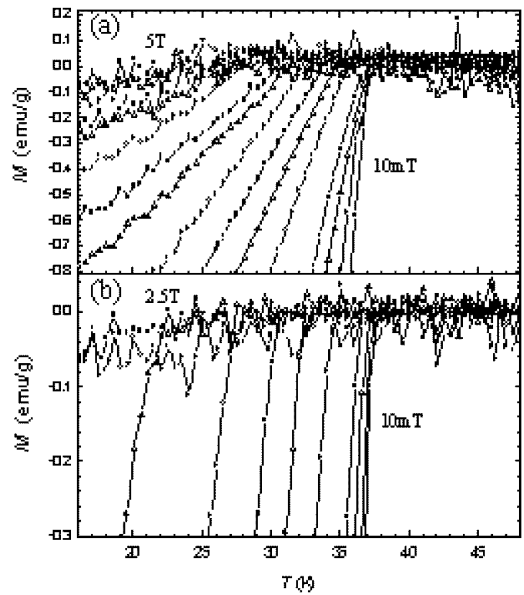}
\end{center}
\vspace*{-0.8cm}
	\caption{(a) and (b) Temperature dependence of the magnetization on zero-field cooling in the several fields.}
	\label{fig:depend}
\end{figure}

\begin{figure}
\begin{center}
\includegraphics[width=12cm,height=12cm,keepaspectratio]{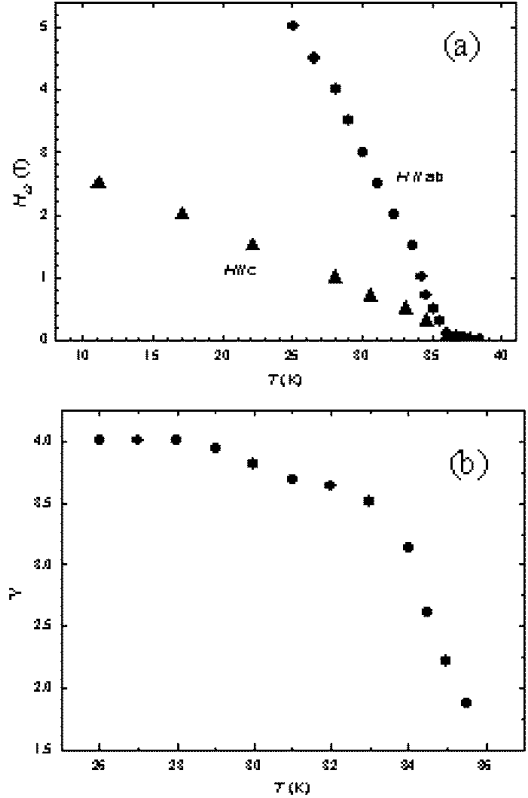}
\end{center}
\vspace*{-0.6cm}
	\caption{(a) Upper critical field for \textit{H}$_{c2}^{ab}$ and \textit{H}$_{c2}^{c}$ determined from the onset of 
the superconductivity in Fig \ref{fig:depend}(a) and (b) as a function of the temperature. 
(b) Temperature dependence of upper critical field anisotropy $\gamma$\,=\,\textit{H}$_{c2}^{ab}$/\textit{H}$_{c2}^{c}$.}
	\label{fig:anisto}
\end{figure}

Figure \ref{fig:depend} shows the temperature dependence of magnetization 
\textit{M}(\textit{T}) curves on warming after field cooling 
the sample for (a) \textit{H}//ab in magnetic field up to 5\,T and (b) 
\textit{H}//c in magnetic field up to 2.5\,T. 
The superconducting transition shifts to lower temperatures as the field increased. 
The superconducting transition in fields is determined by extrapolating the \textit{M}(\textit{T}) 
curve lineally and by finding the crossing point to the horizontal line extended from the normal state. 
From this data, 
Fig. \ref{fig:anisto}(a) shows the upper critical field of MgB$_{2}$ for applied fields 
\textit{H}//ab and \textit{H}//c. \textit{H}$_{c2}^{ab}$ show a non-linear 
temperature dependence
 near \textit{T}c and then rise rapidly at lower temperatures. \textit{H}$_{c2}^{c}$ increases lineally 
with decreasing temperature. 
From Fig \ref{fig:anisto}(a), $\gamma$\,=\,\textit{H}$_{c2}^{ab}$/\textit{H}$_{c2}^{c}$ is found to be temperature dependent, 
and it is shown in Fig \ref{fig:anisto}(b).
It increases from about 1 near \textit{T}c to 4.0 at 25\,K. The extrapolation of \textit{H}$_{c2}^{ab}$ and 
\textit{H}$_{c2}^{c}$ lines to the zero temperature axis yields \textit{H}$_{c2}^{ab}$(0)$\sim$13.6\,T, and
 \textit{H}$_{c2}^{c}$(0)$\sim$3.4\,T with $\gamma$$\sim$4.0.   
These values are similar to the previous 
results obtained from 
magnetic measurements on powder samples~\cite{budko2} or on single crystals~\cite{angst,welp,zehetmayer,sologubenko2},
 but don't agree 
with the reported $\gamma$-values determined from resistivity measurements 
on crystals~\cite{lee,pradhan,kim,xu,simon,lima,perkins} and 
\textit{c}-axis oriented films~\cite{patnaik,jung2,olsson}, which are around 
1.1 to 3, as shown in Table~\ref{tab:table}.

\begin{figure}
\begin{center}
\includegraphics[width=8cm,height=8cm,keepaspectratio]{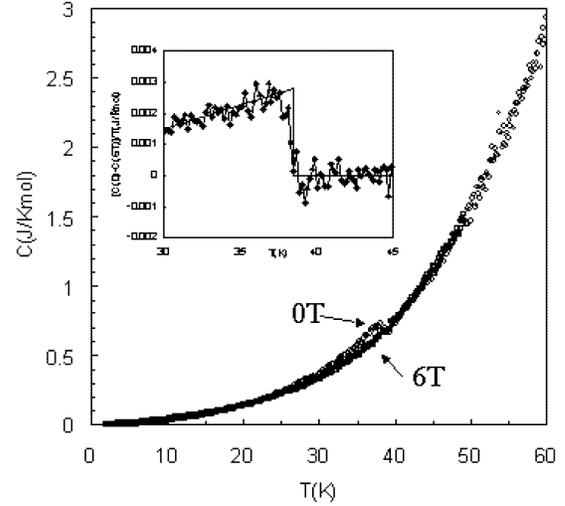}
\end{center}
\vspace*{-0.8cm}
	\caption{Specific heat of MgB$_{2}$ under magnetic field of range 2-60\,K of temperature, 0 and 6\,T. Inset: 
Expansion of \textit{Tc} neighborhood.}
	\label{fig:specific}
\end{figure}

\begin{figure}
\begin{center}
\includegraphics[width=8cm,height=8cm,keepaspectratio,clip]{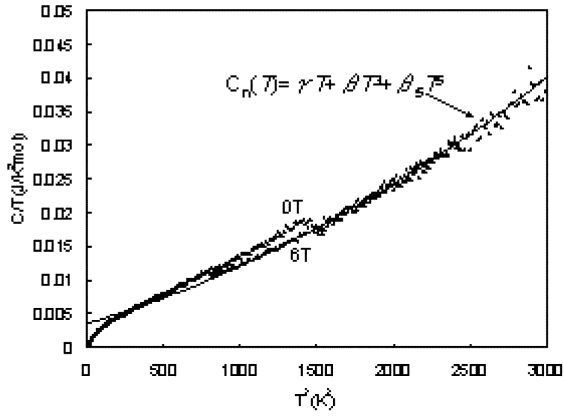}
\end{center}
\vspace*{-0.8cm}
	\caption{\textit{C}(\textit{t})/\textit{T} in magnetic field 0 and 6 \,T plotted as a function of \textit{T}$^{2}$. 
The solid line is a curve by which the fitting is done by 
\textit{C}$_{n}$(\textit{T})\,=\,$\gamma$\textit{T}+$\beta$\textit{T}$^{3}$+$\beta$$_{5}$\textit{T}$^{5}$ and the specific 
heat of normal state is extended to the low temperature.}
	\label{fig:fit}
\end{figure}

\begin{figure}
\begin{center}
\includegraphics[width=8cm,height=8cm,keepaspectratio]{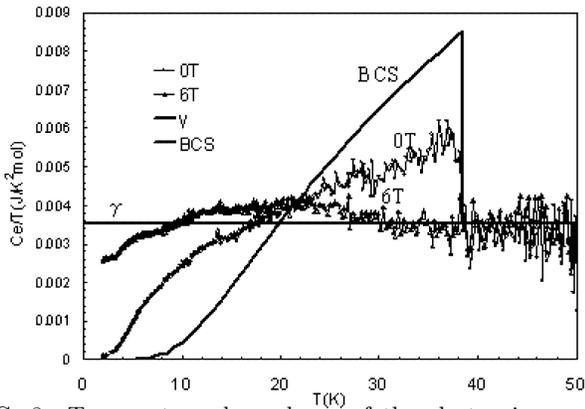}
\end{center}
\vspace*{-1cm}
	\caption{Temperature dependence of the electronic specific heat \textit{C}$_{es}$ in magnetic field 0 and 6\,T. 
The BCS theory was shown in the solid line and the specific heat of normal state was shown in the horizontal dotted
 line.}
	\label{fig:electron}
\end{figure}

The specific heat was measured by a adiabatic 
heat pulse method in the magnetic field of 0 and 6\,T in a temperature range between 2 and 60\,K.
The result of the specific heat as a function of temperature is shown in figure \ref{fig:specific}. The change in the 
specific heat in the \textit{Tc} neighborhood is inserted as an inset in Fig \ref{fig:specific}. The critical temperature 
\textit{Tc} defined by the mid point of the jump in the specific heat is 38.5\,K, and the width of the jump is about 
1\,K. The value $\Delta$\textit{C}(\textit{Tc})/\textit{Tc} of the jump of specific heat is 2.8\,mJ/K$^{2}$\,mol. 
However, the jump of the specific heat was not observed in the magnetic field of 6\,T in this scale. 
It is important to separate electronic specific heat from the entire specific heat to know the excitation
 of the electron system. It is assumed that the entire specific heat is composed of the 
specific heat of the electron system and the lattice system, 
\textit{C}(\textit{T})\,=\,\textit{C}$_{e}$(\textit{T})+\textit{C}$_{ph}$(\textit{T}). The specific heat 
of the lattice system is expressed 
by \textit{C}$_{ph}$(\textit{T})\,=\,$\beta$\textit{T}$^{3}$ within the range of the temperature which is 
sufficiently lower than that 
of the Debye temperature $\theta$$_{D}$. The electron specific heat is assumed to be  
\textit{C}$_{e}$(\textit{T})\,=\,$\gamma$\textit{T} 
in the normal state. The plot of \textit{C}(\textit{T})/\textit{T} vs. \textit{T}$^{2}$ is shown in Fig.\ref{fig:fit}. 
However, 
it is clear that the expression of \textit{C}(\textit{T})/\textit{T}\,=\,$\gamma$+$\beta$\textit{T}$^{2}$ 
in the normal state between 40 and 60\,K 
in 0\,T magnetic field and between 30 and 50\,K in 6\,T magnetic field. 
The higher-order term is then added in the expression to correct 
the specific heat of the lattice: it is approximated by 
\textit{C}$_{ph}$(\textit{T})\,=\,$\beta$\textit{T}$^{3}$\,+\,$\beta$$_{5}$\textit{T}$^{5}$. The electronic 
specific heat was separated
 by extrapolating it to a lower temperature region by using the specific heat 
of the lattice by which the specific heat of the normal state was corrected. 
This fitting gave the result of $\gamma$\,=\,3.51\,mJ/K$^{2}$\,mol, 
$\beta$\,=\,6.76$\times$10$^{-6}$\,J/K$^{4}$\,mol 
and $\beta$$_{5}$\,=\,1.08$\times$10$^{-9}$\,J/K$^{6}$\,mol, 
giving 0.8 to the jump $\Delta$\textit{C}(\textit{Tc})/$\gamma$\textit{Tc} of the specific heat 
normalized by $\gamma$\textit{Tc}, and is different from 1.43 of BCS theory. 
The temperature dependence of the superconducting electronic specific heat \textit{C}$_{es}$(\textit{T})
 which had been obtained by subtracting the specific heat of the lattice from the entire 
specific heat of the superconducting state 
was shown in Fig. \ref{fig:electron}. The temperature dependence of the electronic specific heat of 
the superconducting state in zero magnetic field 
deviates strongly from the specific heat of the BCS theory as seen in Fig. \ref{fig:electron}. It is smaller than the 
value of the BCS theory in 20\,K$\leq$\textit{T}$\leq$\textit{Tc}. 
But it is larger than the value of the BCS theory in \textit{T}$\leq$20\,K. A gradual jump of the 
specific heat, which extended in a wide 
range of temperature 20-28\,K under the magnetic field of 6Tesla was observed. The critical 
temperature in this magnetic field 
agrees roughly to the one obtained by the experimental result of the upper critical magnetic 
field measurement~\cite{xu}. 

\begin{figure}[th]
\begin{center}
\includegraphics[width=8cm,height=8cm,keepaspectratio,clip]{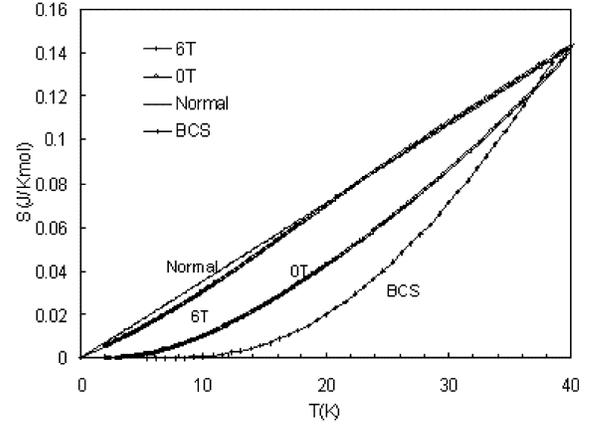}
\end{center}
\vspace*{-0.9cm}
	\caption{Entropy as function of temperature in magnetic field of 0 and 6\,T. The entropy of normal state and the 
BCS theory was shown.}
	\label{fig:entropy}
\end{figure}

The entropy in the superconducting state can be obtained by using 
\textit{S}(\textit{T})\,=\,$\int$\textit{C}$_{e}$(\textit{T})/\textit{TdT} under 
the condition that the entropy of superconducting state and 
normal state should be equal at \textit{Tc}, i.e., \textit{S}$_{s}$(\textit{Tc})\,=\,\textit{S}$_{n}$(\textit{Tc}). 
The temperature dependencies of the entropy in both states are 
shown in Fig. \ref{fig:entropy}.  
Furthermore, the critical magnetic field \textit{Hc}(\textit{T}) was obtained from the entropy of superconducting 
state and normal 
state by using the expression 
$\int${\textit{S}$_{n}$(\textit{T})\,-\,\textit{S}$_{s}$(\textit{T})}\textit{dT}\,=\,\textit{H}$_{c}$(\textit{T})$^{2}$/8$\pi$.

\begin{figure}
\begin{center}
\includegraphics[width=10cm,height=10cm,keepaspectratio,clip]{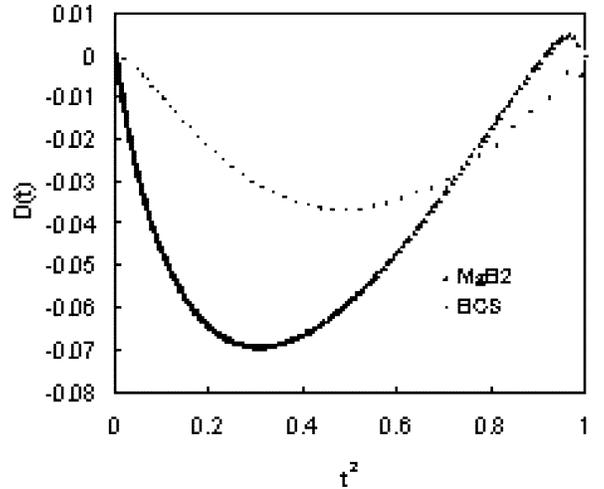}
\end{center}
\vspace*{-1cm}
	\caption{Temperature dependence of the deviation function \textit{D}(\textit{t}): 
\textit{D}(\textit{t})\,=\,\textit{H}$_{c}$(\textit{t})/\textit{H}$_{c}$(\textit{0})-(1-\textit{t}$^{2}$), 
\textit{t}\,=\,\textit{T}/\textit{Tc}.}
	\label{fig:D(t)}
\end{figure}

In order to make the difference from the BCS theory clear, 
the deviation function 
\textit{D}(\textit{t})\,=\,\textit{H}$_{c}$(\textit{t})/\textit{H}$_{c}$(0)\,-\,(1-\textit{t}$^{2}$) 
as shown in Fig. \ref{fig:D(t)} where \textit{t}\,=\,\textit{T}/\textit{Tc}. The BCS theory assumes 
the weak electron-phonon interaction, the isotropic Fermi surface and the isotropic interaction. 
The difference between the superconducting electronic specific heat and 
the one of the BCS theory shown in Figure \ref{fig:D(t)}
 may be  caused by the strong coupling effect, and/or the anisotropy of the Fermi surface and the anisotropy 
of the interaction. However, the effect of the electron-phonon coupling and anisotropy cause 
the opposite effect in the normalized jump of the specific heat $\Delta$\textit{C}(\textit{Tc})/$\gamma$\textit{Tc} 
and the deviation function \textit{D}(\textit{t}). 
That is, when the coupling becomes stronger, the value of $\Delta$\textit{C}(\textit{Tc})/$\gamma$\textit{Tc} 
grows more than the BCS value of 1.43, and 
the deviation function \textit{D}(\textit{t}) shifts from the BCS curve to a positive direction. On the other hand, 
when the anisotropy becomes , the value of $\Delta$\textit{C}(\textit{Tc})/$\gamma$\textit{Tc} 
becomes smaller than the values of BCS, 
and \textit{D}(\textit{t}) is changed in a negative direction~\cite{clem,padamsee,gubser}. Value 0.8 of the 
jump of the normalized specific heat 
was smaller than the value of BCS and show the deviation function \textit{D}(\textit{t}) in Fig. \ref{fig:D(t)} 
which appeared in more 
negative direction than the BCS curve.
 
According to the results shown above the superconducting state of MgB$_{2}$ 
exhibits rather strong anisotropy of order 4 at low temperature.
 This result agrees with the result of the specific 
heat measurement in other groups~\cite{bouquet}. The experimental result predicting the existence of 
two gaps has been reported in the result of tunneling microscopy~\cite{giubileo}. 
Models of the two-zone or the multi-zone where the anisotropy is treated by dividing the Fermi surface, 
when the anisotropy of both the Fermi surface and the coupling becomes stronger, might be a good model. 
The anisotropy effect in the framework of the two-zone model was discussed~\cite{entel}. In that case, the value of the 
energy gap is different in each zone. Therefore, the specific heat would have the jump at a different temperature 
to which the energy gap of each zone opens. Though the temperature dependency of specific heat changes 
largely by the inter-zone scattering of conduction electrons, it differs largely from specific 
heat of the single-zone (single-band) isotropic superconductor. Part of the large electronic 
specific heat \textit{C}$_{es}$ at the low temperature in Fig. \ref{fig:electron} depends on 
the existence of a small energy gap 
it to be possible to excite even at the low temperature.

In summary, we reported on the results of measurement and the analysis of the specific heat of polycrystalline MgB$_{2}$ 
and the magnetic properties of single crystalline MgB$_{2}$. 
The behavior of the superconducting electronic specific heat and the deviation function indicate 
that the superconducting state of this MgB$_{2}$ is a strong anisotropy superconducting state, 
which is different from the BCS theory. From the magnetization measurement on single crystals, the upper critical field 
anisotropy ratio $\gamma$\,=\,\textit{H}$_{c2}^{ab}$/\textit{H}$_{c2}^{c}$ is found to be increased 
from about 1 near \textit{Tc} to 4.0 at 25\,K.

We would like to thank K. Yamawaki, K. Noda, K. Ishikawa, H. Saito, Y. Takano, and K. Ohshima 
for useful discussions.

\bibliographystyle{apsrev}

\bibliography{refer}
\end{document}